\documentclass[a4paper,11pt]{article}

\usepackage{jinstpub}

\usepackage{lineno}
\usepackage[separate-uncertainty = true]{siunitx}
\usepackage{isotope}
\usepackage{fancyref}
\usepackage{comment}

\usepackage{placeins}
\usepackage{csquotes}

\title{Performance of neutron-irradiated 4H-Silicon Carbide diodes subjected to Alpha radiation}

\author[a,1]{P. Gaggl,\note{Corresponding author.}}
\author[a]{A. Gsponer,}
\author[a]{R. Thalmeier,}
\author[a]{S. Waid,}
\author[b]{G. Pellegrini,}
\author[b]{P. Godignon,}
\author[b]{J.M. Rafí,}
\author[a]{T. Bergauer}

\affiliation[a]{Institute of High Energy Physics of the Austrian Academy of Sciences,\\Nikolsdorfer Gasse 18, Vienna, 1050, Austria}
\affiliation[b]{IMB-CNM-CSIC, Campus UAB, 08193 Bellaterra, Barcelona, Spain}

\emailAdd{Philipp.Gaggl@oeaw.ac.at}

\abstract{
The unique electrical and material properties of 4H-silicon-carbide (4H-SiC) make it a promising candidate material for high rate particle detectors. 
In contrast to the ubiquitously used silicon (Si), 4H-SiC offers a higher carrier saturation velocity and larger breakdown voltage, enabling a high intrinsic time resolution and mitigating pile-up effects. 
Additionally, as radiation hardness requirements grow more demanding in the context of future high luminosity high energy physics experiments, wide-bandgap materials such as 4H-SiC could offer better performance due to low dark currents and higher atomic displacement thresholds.
In this work, the detector performance of 50~\si{\micro\metre} thick 4H-SiC p-in-n planar pad sensors was investigated at room temperature, using an \textsuperscript{241}Am alpha source at reverse biases of up to 1100~\si{\volt}.
Samples subjected to neutron irradiation with fluences of up to $1\times 10^{16}~\text{n}_\text{eq}/\si{\centi\metre\squared}$ were included in the study in order to quantify the radiation hardness properties of 4H-SiC.
A calibration of the absolute number of collected charges was performed using a GATE simulation.
The obtained results are compared to previously performed UV transient current technique (TCT) studies.
Samples exhibit a drop in charge collection efficiency (CCE) with increasing irradiation fluence, partially compensated at high reverse bias voltages far above full depletion voltage. 
At fluences of $5\times 10^{14}~\text{n}_\text{eq}/\si{\centi\metre\squared}$ and $1\times 10^{15}~\text{n}_\text{eq}/\si{\centi\metre\squared}$, CCEs of \SI{64}{\percent} and \SI{51}{\percent} are obtained, decreasing to \SI{15}{\percent} at $5\times 10^{15}~\text{n}_\text{eq}/\si{\centi\metre\squared}$.
A plateau of the collected charges is observed in accordance with the depletion of the volume the alpha particles penetrate for an unirradiated reference detector.
For the neutron-irradiated samples, such a plateau only becomes apparent at higher reverse bias, roughly \SI{600}{\volt} and \SI{900}{\volt} for neutron fluences of $5\times 10^{14}~\text{n}_\text{eq}/\si{\centi\metre\squared}$ and $1\times 10^{15}~\text{n}_\text{eq}/\si{\centi\metre\squared}$.
For the highest investigated fluence, CCE behaves almost linearly with increasing reverse bias.
Compared to UV-TCT measurements, the reverse bias required to deplete a sensitive volume covering full energy deposition is lower, due to the small penetration depth of the alpha particles.
At the highest reverse bias, the measured CCE values agree well with earlier UV-TCT studies, with discrepancies between \SI{1}{\percent} and \SI{5}{\percent}.
}


\proceeding{23$^{\text{rd}}$ International Workshop on radiation Imaging Detectors\\26-30 June 2022, Riva del Garda, Italy}

\begin{document}
\maketitle
\flushbottom

\section{Introduction}\label{sec:intro}
The lion's share of present-day solid state particle detectors are silicon (Si) based, owing to its technical maturity and favourable electrical properties.
The ubiquitous use of silicon electronics brings forth an in-depth understanding of manufacturing processes as well as significant research and development efforts, factors the field of high energy physics profits extensively from.

Recently, silicon carbide (SiC) has been garnering significant attention as a detector material, with many different applications in neutron~\cite{neutron_detection_2, neutron_detection}, alpha-particle~\cite{alpha_detection}, and high energy particle detection~\cite{paper_HEPHY, review_sic, gsponer_rd50}.
The interest in silicon carbide, as well as other wide-bandgap detector materials, has been mainly fueled by the radiation hardness requirements of future high luminosity experiments~\cite{LHC_upgrade}, where silicon detectors might prove insufficient~\cite{comparison_radiation_hard_devices, rd50_sic_paper}.
Currently, different manufacturing processes~\cite{paper_DOFZ_Si} and detector geometries~\cite{paper_3d_detectors} are being investigated for silicon detectors~\cite{paper_LHC_upgrade_proposals}.
However, wide-bandgap materials might offer enhanced radiation hardness even with current detector designs.
Amongst a high amount of various polytypes, defined by the individual crystal structure and stacking sequence~\cite{book_fundamentals_of_SiC}, 4H-SiC arguably exhibits the most favorable material properties.
Its high displacement threshold energy suppresses defect creation~\cite{book_fundamentals_of_SiC}.
The large band gap of 4H-SiC results in much lower leakage currents when compared to Si, reducing noise and allowing for operation at high temperatures~\cite{RD50_high_temperature}.
Furthermore, 4H-SiC is insensitive to the visible spectrum.

A charge carrier saturation velocity about twice that of Si provides faster detector signals, improving time resolution and mitigating particle pile-up in high-rate applications~\cite{book_fundamentals_of_SiC}.
However, the ionization energy of SiC is about two times larger than for Si~\cite{rd50_sic_paper, E_ion_high}.
In combination with only relatively thin (< $\SI{100}{\micro\meter}$) SiC detectors being available due to cost and manufacturing limitations, high signal-to-noise detection of single minimum ionizing particles (MIPs) can be challenging~\cite{paper_HEPHY}.

\section{Materials and Methods}\label{sec:materials_and_methods}
The measurements presented in this report are a continuation of charge collection efficiency (CCE) studies conducted on the same neutron-irradiated samples using a UV transient current technique (TCT) laser setup, described in~\cite{Gaggl_2022}.

\subsection{4H-SiC Samples}
Measurements were conducted on an unirradiated and neutron-irradiated planar 3$\times$3~\si{\milli\metre\squared} p-on-n diodes, manufactured and developed at IMB-CNM-CSIC~\cite{URL_CNM}.
The active region consists of a 50~\si{\micro\metre} n-doped epitaxially grown active layer, with a resistivity of 20~\si{\ohm\centi\metre} and a full depletion voltage of 296~\si{\volt} in the unirradiated case~\cite{paper_HEPHY}.
Further information about the used samples, their geometric structure and composition, as well as the contact window, can be found in~\cite{Gaggl_2022, paper_samples, paper_samples_contact_window}. 
The 4H-SiC samples were mounted and wire-bonded on a UCSC single channel board~\cite{URL_UCSC} (also known as \enquote{LGAD Board} due to its frequent usage in conjunction with LGAD detectors), which features a high-bandwidth transimpedance-amplifier (TIA).
A detailed description on modifications to the board and the detector fixation can be found in~\cite{paper_HEPHY, Gaggl_2022}.

\subsection{Neutron Irradiation}\label{sec:neutron_irradiation}
4H-SiC samples have been neutron-irradiated at four different fluences at the TRIGA Mark II nuclear research reactor located at the Atominstitut in Vienna~\cite{TRIGA_source}. The resulting fluences ($5 \times 10^{14}$, $1 \times 10^{15}$, $5 \times 10^{15}$ and $1 \times 10^{16}~\text{n}_\text{eq}/\si{\centi\metre\squared}$, given in 1~\si{\mega\electronvolt} neutron equivalent according to NIEL hypothesis~\cite{NIEL}) are a direct result of various exposition times, summarized in~\cite{Gaggl_2022}.
Further information on the research reactor and irradiation position, including a description of the local neutron spectrum and corresponding correlation factors are given in~\cite{TRIGA_source} and~\cite{TRIGA_neutron_fluences}.

\subsection{Alpha Measurement}
\label{sec:alpha_GATE}
\begin{figure}[htp]
\centering
\fbox{\includegraphics[width=0.45\textwidth]{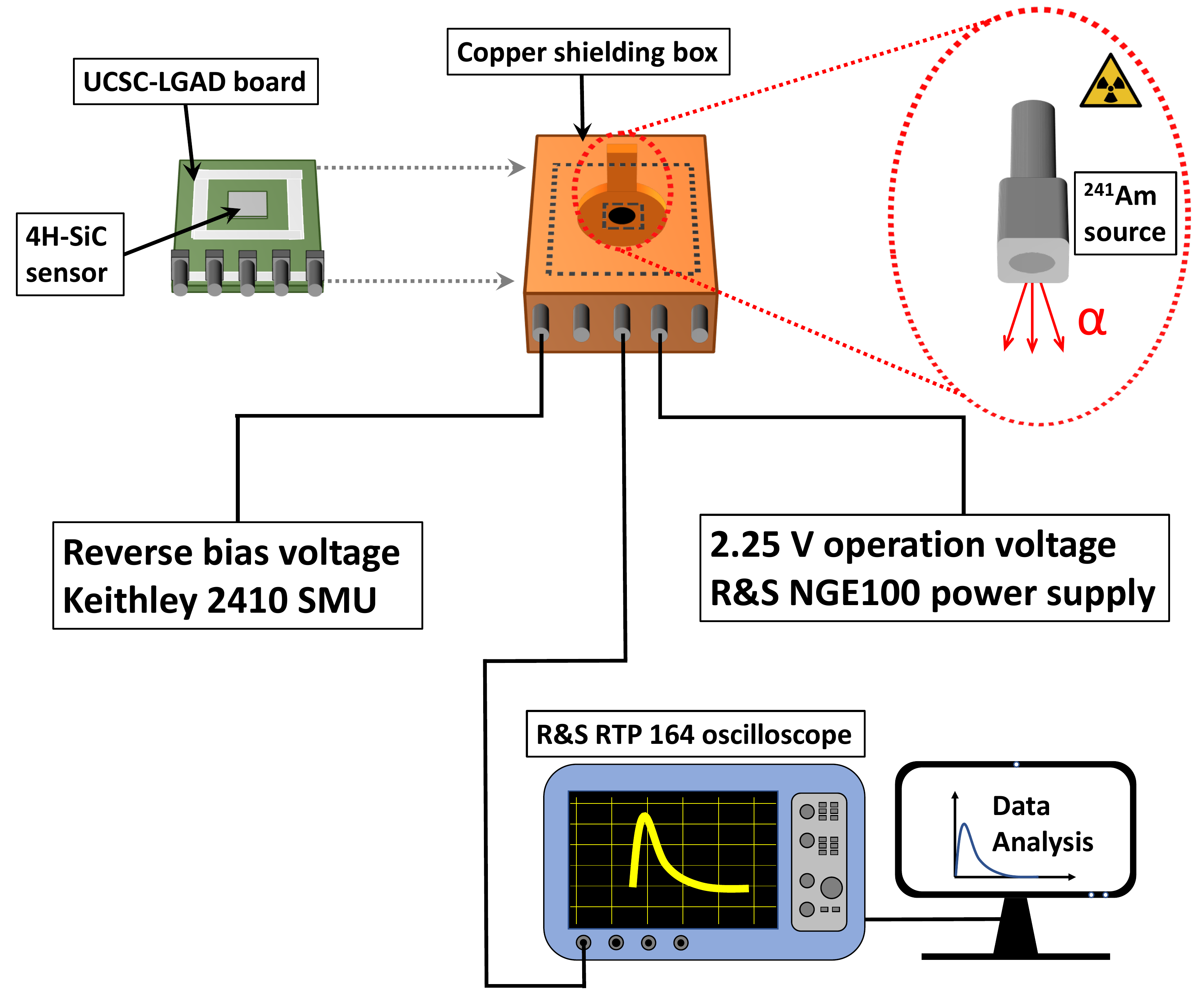}}
\vspace{-0.25cm}
\caption{Schematic overview of the experimental setup}
\label{fig:Alpha_source}
\end{figure}

A 9.5~\si{\kilo\becquerel} \textsuperscript{241}Am source from Eckert~\&~Ziegler was used for alpha measurements.
The isotope is concentrated on a 8~\si{\milli\metre} diameter disk with a thickness of about 1~\si{\micro\metre}.
The emitter is covered by a \SI{1.4}{\micro \meter} protective gold layer and located within a steel collimator.
A sketch of the setup is shown in \autoref{fig:Alpha_source}.
The UCSC LGAD boards, on which the 4H-SiC samples are mounted, were inserted into a custom-made copper box.
In addition to providing RF shielding, the box includes a milled slot for the \textsuperscript{241}Am source.
This allows for a constant distance between the source and detector across different measurements, minimizing the variability of scattering in air.
The resulting source-detector distance is \SI{8.5(1)}{\milli \meter}.
Although a shorter distance would be preferable, the necessary clearance required for the bonding wires acted as a limiting factor.

For applying the reverse bias, a Keithley 2410 source meter was used, while the \SI{2.25}{\volt} supply voltage of the TIA was generated by a Rohde\&Schwarz NGE100 power supply.
The signals were digitized by a Rhode\&Schwarz RTP 164 oscilloscope with \SI{40}{\giga Sa \per \second} and \SI{16}{\giga \hertz} analog bandwidth.
Because of the high bandwidth of the used TIA and the small detector thickness, the obtained signals are typically shorter than \SI{2}{\nano \second}, necessitating a fast oscilloscope to properly resolve the signal shape.
The oscilloscope's internal trigger was set to \SI{2}{\milli \volt}, which is slightly above the noise floor of the setup. 
The analysis was performed offline and is described in \autoref{sec:analysis}. It needs to be highlighted, that all measurements were performed at room temperature.

\subsection{Calibration using GATE Simulations} \label{sec:GATE}
\begin{figure}[htp]
\centering
\includegraphics[width=0.55\textwidth]{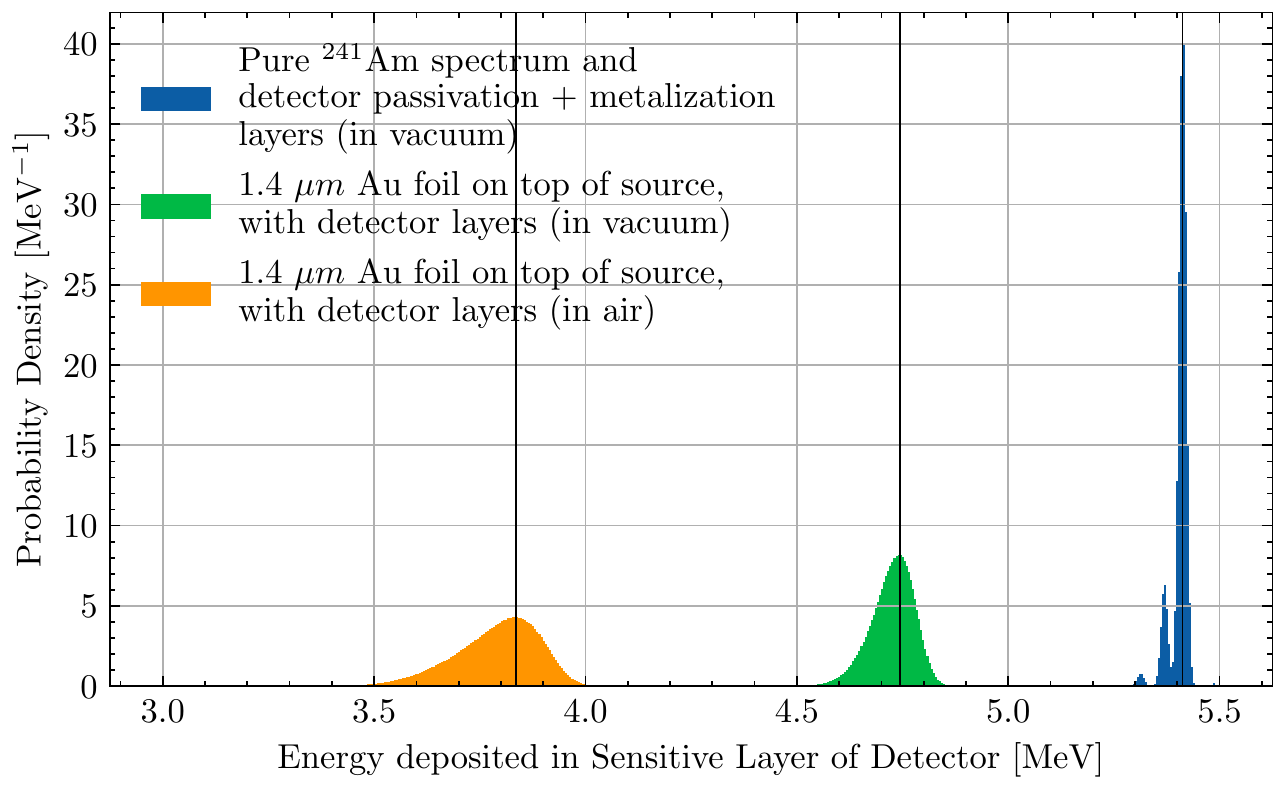}
\vspace{-0.25cm}
\caption{Distribution of energy deposited by alpha particles in the sensitive epitaxial volume for different geometries. The black vertical lines indicate the most probable values. In vacuum and with a pure \textsuperscript{241}Am spectrum (no protective gold foil), the energy straggling in the passivation and metalization layers is limited (blue data) and the three decay energies of \textsuperscript{241}Am can still be separated. A considerably broader energy spectrum is observed when the gold foil covering the source is introduced (green). If the air gap (\SI{8.5}{\milli \meter}) is also added to the simulation, the distribution grows even broader.}
\label{fig:Alpha_Simulations}
\end{figure}

In order to use the energy deposited by \textsuperscript{241}Am alpha particles as a calibration reference, the energy distribution arriving at the detector needs to be characterized.
There are three main decay energies for~\textsuperscript{241}Am: \SI{5485.56}{\kilo\electronvolt} (84.80~\si{\percent}), \SI{5442.80}{\kilo\electronvolt} (13.10~\si{\percent}) and \SI{5388.00}{\kilo\electronvolt} (1.66~\si{\percent})~\cite{URL_241Am_decay_energies}.
However, due to the material between the source and the sensitive layer of the detector, some energy loss and straggling is expected, especially within the gold passivation layer of the alpha source and the air gap between source and detector.
In order to quantify this energy loss, a GATE~\cite{GATE_publication} simulation was performed.
The simulation took into account the source geometry (gold protection layer and collimator), as well as passivation (\SI{100}{\nano \meter} Si\textsubscript{3}N\textsubscript{4} + \SI{50}{\nano \meter} SiO\textsubscript{2}) and metalization (\SI{235}{\nano \meter} Ti/Al/Ti/W stack) covering the detector~\cite{paper_samples}.
The simulated energy deposition in the detector for three different cases is shown in \autoref{fig:Alpha_Simulations}.

The gold protection layer and air gap are the major contributors to the energy straggling, absorbing \SI{0.66}{\mega\electronvolt} and \SI{0.84}{\mega\electronvolt} when considering the most probable value (MPV).
In comparison, the energy loss in the passivation and metalization layers is insignificant~(\SI{0.07}{\mega\electronvolt}).
For the complete geometry (orange data set in \autoref{fig:Alpha_Simulations}), a deposited energy between \SI{3.5}{\mega\electronvolt} and \SI{4.0}{\mega\electronvolt} is expected in the sensitive volume of the detector, with a MPV of~\SI{3.83}{\mega\electronvolt}.
The simulated penetration depth ranges from \SI{12}{\micro \meter} to \SI{17}{\micro \meter}.
As this is smaller than the thickness of the sensitive layer (\SI{50}{\micro \meter}), the alpha particle kinetic energy is fully absorbed.

\FloatBarrier
\section{Measurements}\label{sec:measurements}
In total, five 4H-SiC samples were investigated (four neutron-irradiated ones according to \autoref{sec:neutron_irradiation} and one unirradiated reference).
For the sample subjected to the highest neutron fluence ($1\times 10^{16}~\text{n}_\text{eq}/\si{\centi\metre\squared}$), no significant signals could be obtained at reverse bias voltages lower than \SI{750}{\volt}.
Therefore, this sample was excluded from the analysis. A reverse bias voltage scan ranging from \SI{50}{\volt} to \SI{1100}{\volt} with a step size of 50~\si{\volt} was performed at room temperature.
For each step, \num{30000} signals were recorded.
In order to only perform measurements in a stable state, a settling time of at least \SI{30}{\second} was used between changes in the bias voltage.

In case of the sample irradiated to $5\times 10^{15}~\text{n}_\text{eq}/\si{\centi\metre\squared}$, data could only be acquired up to a reverse bias of 1000~\si{\volt} due to a high-voltage breakdown within the readout electronics.
The choice of applying reverse biases up to more than three times the full depletion voltage is justified by the high charge carrier saturation velocity of 4H-SiC, exploitable only at sufficiently high electric fields.
Furthermore, previous UV-TCT measurements implied a partial compensation of the irradiation-induced signal loss at higher reverse bias~\cite{Gaggl_2022}.
When possible, each voltage scan was measured consecutively within a single day and without interruption to reduce the influence of changing external conditions such as temperature, humidity or air pressure.

IV measurements on the same samples were performed beforehand and are given in~\cite{Gaggl_2022}.
For all samples, dark current levels remained within the \si{\nano\ampere}-range.
However, these measurements were done with the pad diodes connected to the UCSC LGAD boards, which are equipped with a HV filtering network of capacitors, also contributing to the measured current. 
IV-characteristics from similar 4H-SiC devices, measured by directly probing the detector, report currents in the sub \si{\pico\ampere}-range for the unirradiated case, only increasing up to 10~\si{\pico\ampere} after neutron irradiation to $1\times 10^{16}~\text{n}_\text{eq}/\si{\centi\metre\squared}$~\cite{paper_samples}.

\section{Analysis}\label{sec:analysis}
For data analysis, an open-source automated pulse finding and analysis software based on Python has been used \cite{URL_git}.
A signal threshold identical to the hardware trigger threshold was applied for pulse finding, requiring a minimum of ten consecutive samples (corresponding to 250~\si{\pico\second}) above the threshold.
To calculate the time-over-threshold (ToT) for each signal, a constant fraction discrimination (CFD) at 25\% of the signal amplitude was used.
The noise before each pulse was fitted with a Gaussian to estimate the root mean square (RMS) noise originating from the read-out electronics.
The signal area was calculated by integrating the signal above the estimated RMS noise level.
This area is directly proportional to the collected charge.
In order to normalize the signal area to a number of collected charges, the most probable energy value as simulated in \autoref{sec:GATE} was used.
The charge collection efficiencies (CCEs) of the irradiated sensors were computed by comparing their most probable collected charge to that of an unirradiated sample.

\section{Results}\label{sec:results_and_discussions}

\begin{figure}[ht]
\centering
\includegraphics[width=0.6\textwidth]{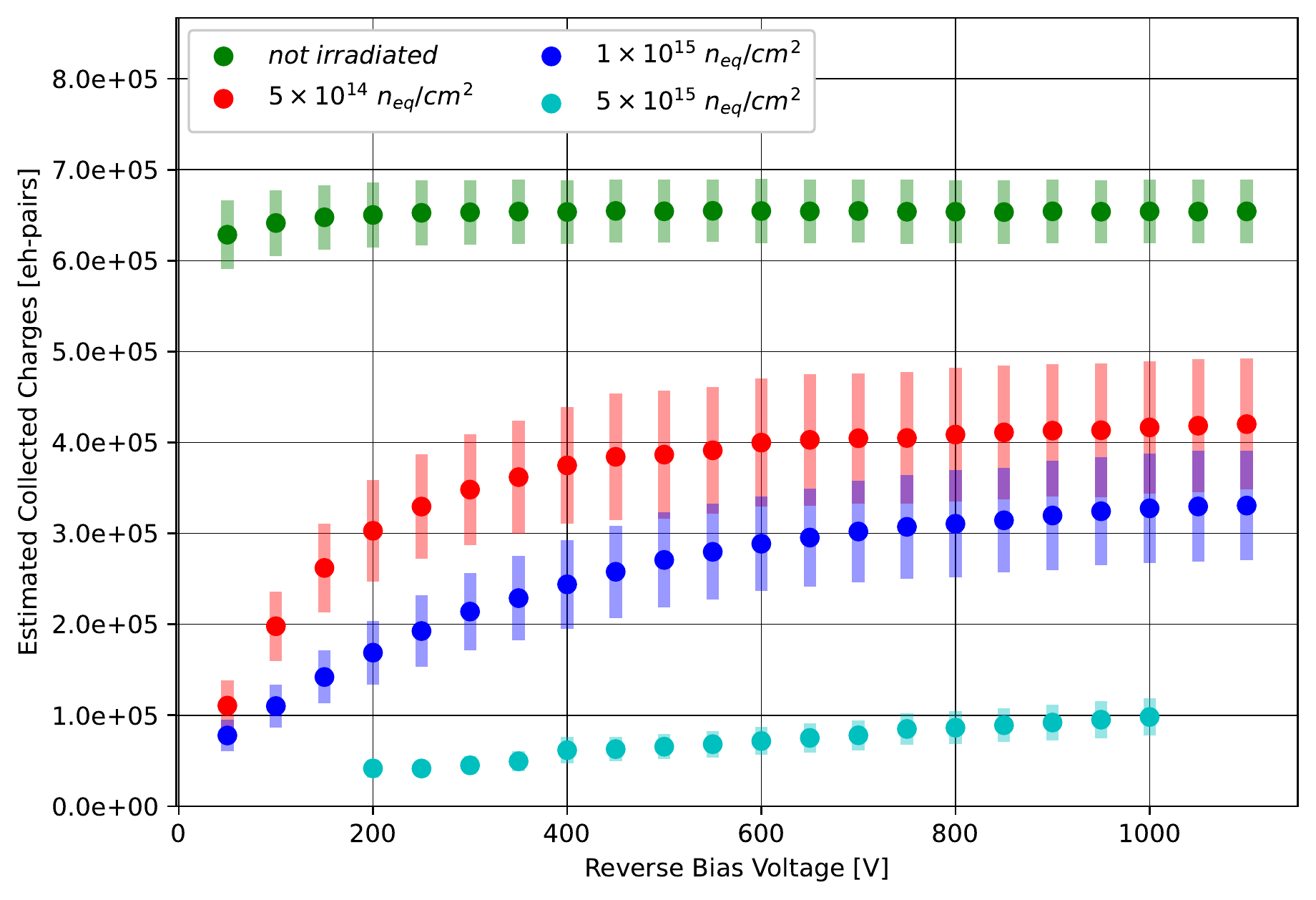}
\vspace{-0.25cm}
\caption{Integrated pulse area for 4H-SiC samples subjected to different neutron irradiation fluences. The data is scaled to the expected amount of electron-hole pairs for a \SI{3.83}{\mega\electronvolt} energy deposition.}
\label{fig:results_total_charges}
\end{figure}

The resulting estimated collected charge for all measured samples is shown in \autoref{fig:results_total_charges}.
After a reverse bias of roughly~\SI{200}{\volt}, the charge collected by the unirradiated sample remains constant.
The discrepancy compared to the measured full depletion voltage of 296~\si{\volt} \cite{paper_HEPHY} is a direct consequence of the low penetration depth of alpha particles, requiring a lower reverse bias voltage to cover the entire volume of energy deposition.
A saturation of collected charges for samples subjected to neutron irradiation is not reached up to the maximum measured reverse bias voltage of 1100~\si{\volt}.
However, for lower irradiation fluences of $5\times 10^{14}~\text{n}_\text{eq}/\si{\centi\metre\squared}$ and $1\times 10^{15}~\text{n}_\text{eq}/\si{\centi\metre\squared}$, saturation is imminent at higher reverse bias voltages of roughly \SI{600}{\volt} and \SI{900}{\volt}.
No indication of saturation is observed for the sample irradiated to $5\times 10^{15}~\text{n}_\text{eq}/\si{\centi\metre\squared}$.
This indicates that a higher reverse bias voltage (equal to a larger drift velocity) can partially compensate an increased defect density, as the probability of charge trapping is proportional to the drift time of the charge carriers.
This is also supported by former UV-TCT measurements~\cite{Gaggl_2022}.

\begin{figure}[htp]
\centering
\includegraphics[width=0.55\textwidth]{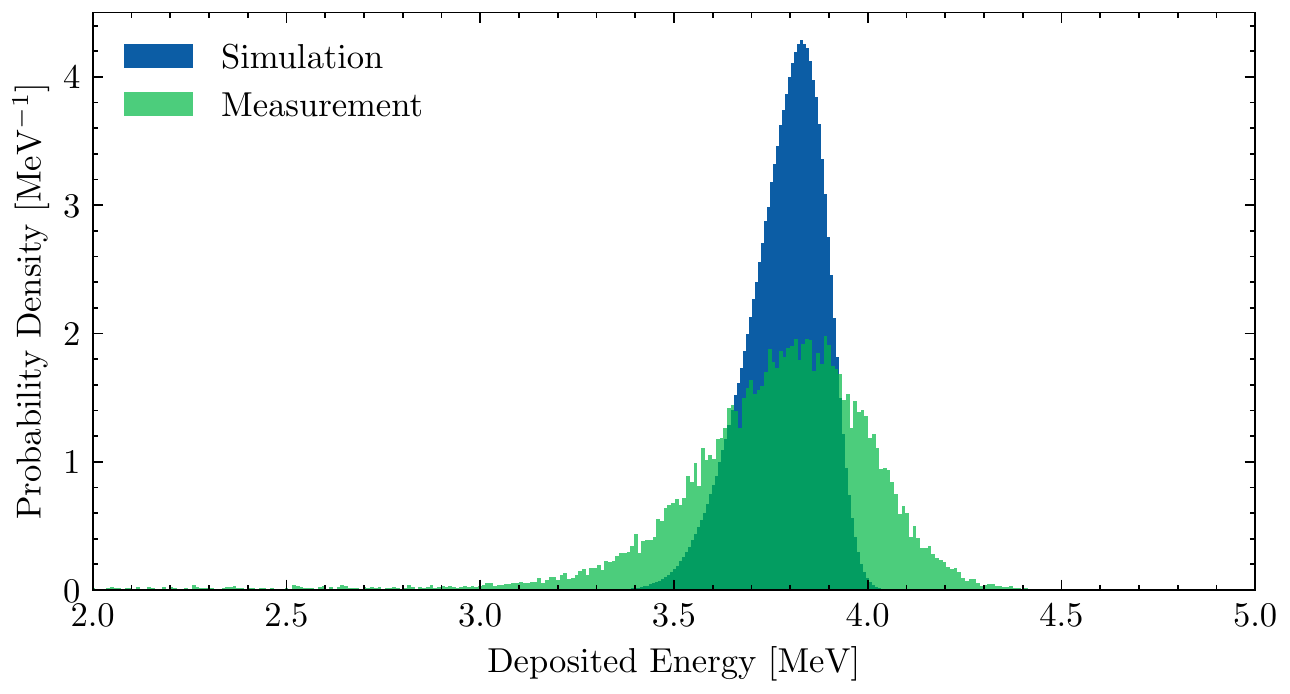}
\vspace{-0.25cm}
\caption{Comparison of the integrated pulse area from the detector (green) and the simulated energy distribution (blue) for an unirradiated sample. The measured signal area has been normalized to match the MPV of the simulated energy distribution (\SI{3.83}{\mega\electronvolt}).}
\label{fig:Alpha_vs_GATE}
\end{figure}

\autoref{fig:Alpha_vs_GATE} shows a histogram comparing the measured charges collected by the unirradiated reference detector with simulation results (see \autoref{sec:GATE}).
The width of the energy distribution is almost twice as broad for the measured data than expected.
A possible explanation for this is the noise introduced by the read-out electronics, but this is still a topic of ongoing investigations.
For both data sets, a slightly skewed distribution can be observed, originating from the three different decay energies of~\textsuperscript{241}Am.

\begin{figure}[ht]
\centering
\includegraphics[width=0.55\textwidth]{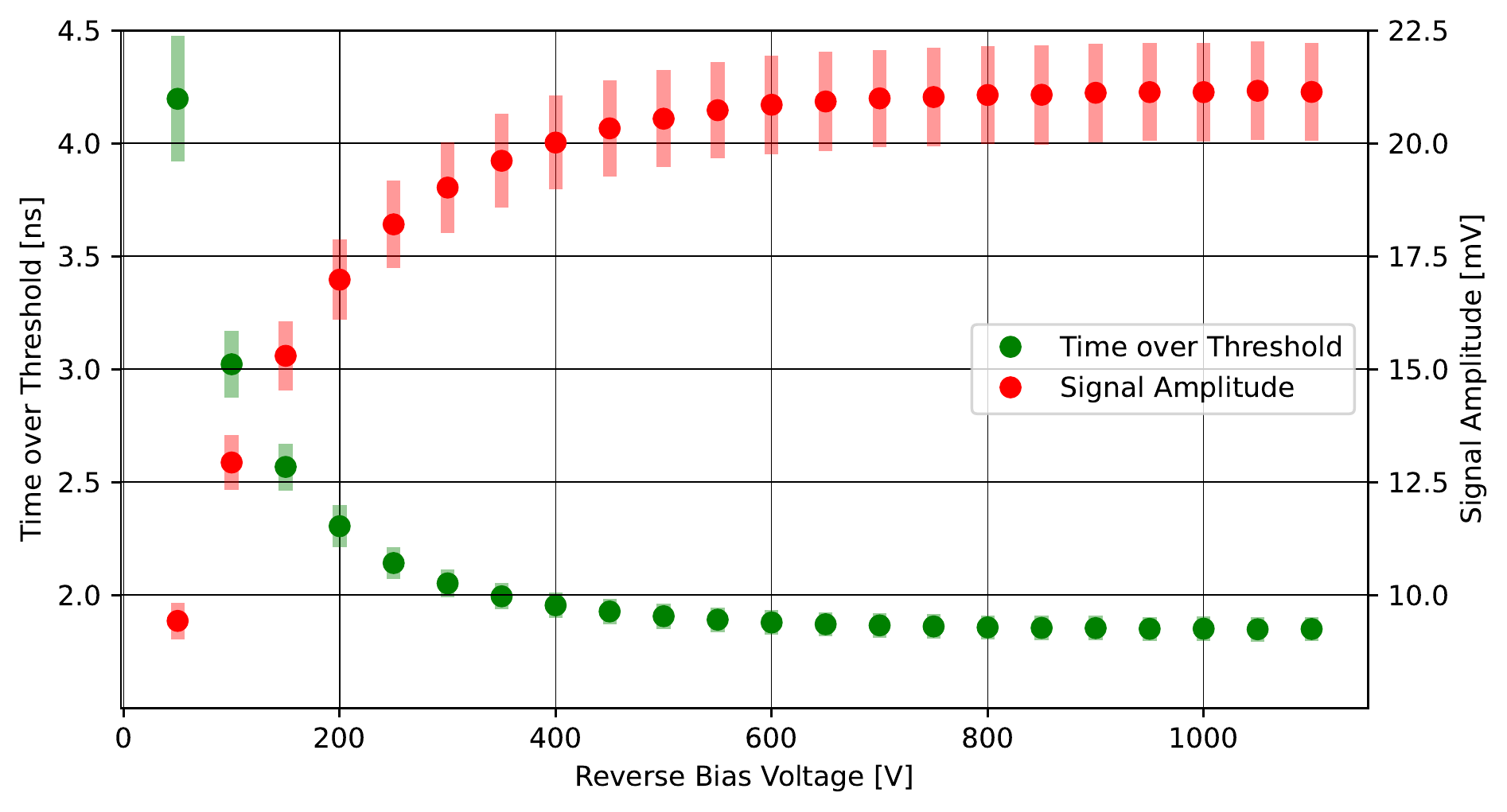}
\vspace{-0.25cm}
\caption{Time over threshold (ToT, green) and signal amplitude (red) as a function of the reverse bias voltage for the unirradiated reference sample.}
\label{fig:tot_max}
\end{figure}

The ToT and signal amplitude saturate only at a much higher reverse bias, as illustrated in \autoref{fig:tot_max} for the unirradiated reference sensor.
As the total collected charge remains constant at higher bias voltages (\SI{100}{\percent} charge collection efficiency), the ToT decreases inversely proportional to the signal amplitude.
After the saturation velocity of the charge carriers is reached, the ToT does not decrease any further.

\begin{figure}[ht]
\centering
\includegraphics[width=0.65\textwidth]{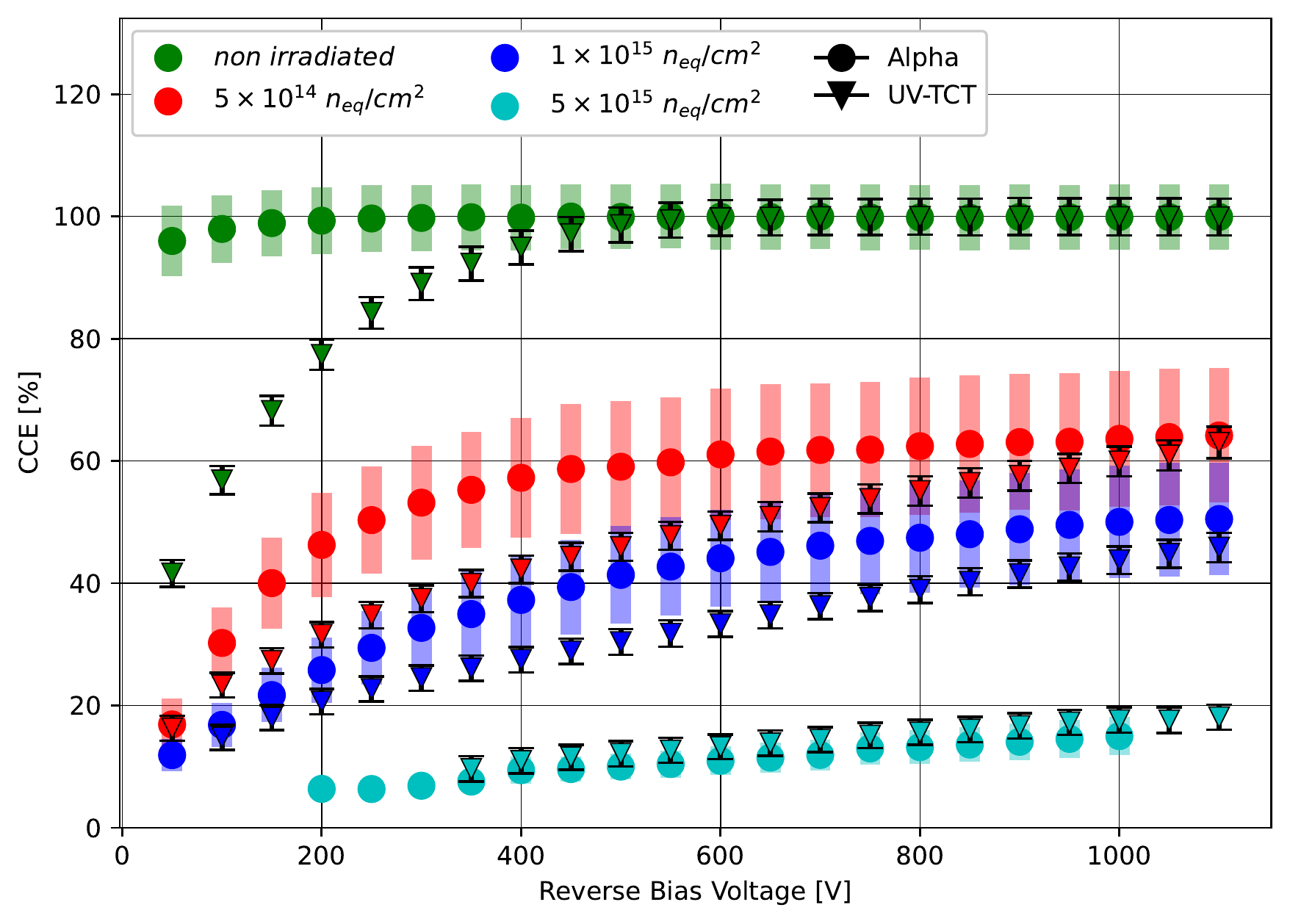}
\vspace{-0.25cm}
\caption{Measured charge collection efficiency (CCE) for samples subjected to neutron irradiation at various fluences obtained using \textsuperscript{241}Am alpha particles and UV laser pulses. A CCE of 100\% corresponds to the respective maximum collected charge of the unirradiated sample.}
\label{fig:results_CCE_comp}
\end{figure}

The charge collection efficiency (CCE) for all samples is shown in \autoref{fig:results_CCE_comp}.
A decrease in the CCE with increasing neutron irradiation fluence is apparent, resulting from the increasing number of generation-recombination centers with irradiation~\cite{paper_samples, paper_neutron, paper_neutron_2}.
Regarding fluences of~$5\times 10^{14}~\text{n}_\text{eq}/\si{\centi\metre\squared}$ and $1\times 10^{15}~\text{n}_\text{eq}/\si{\centi\metre\squared}$, CCEs of \SI{64}{\percent} and \SI{51}{\percent} can still be obtained.
However, CCE decreases drastically for the sample subjected to $5\times 10^{15}~\text{n}_\text{eq}/\si{\centi\metre\squared}$, reaching a maximum of \SI{15}{\percent} at \SI{1000}{\volt}.
An increase in the CCE would still be expected at higher bias voltages.
However, the setup is limited by high-voltage breakdowns of the UCSC LGAD boards.

\autoref{fig:results_CCE_comp} also shows a comparison to former UV-TCT measurements (at 370~\si{\nano\metre}~\cite{Gaggl_2022}) performed on the same samples.
For both data sets, the CCE has been estimated by assuming \SI{100}{\percent} efficiency for the unirradiated reference sample at highest reverse bias.
The previously mentioned early saturation in collected charges under alpha irradiation is apparent in comparison to the UV-TCT data.
For the UV-TCT measurements, the collected charge saturates in accordance with the full depletion voltage, owing to the uniform charge deposition throughout the sensitive detector volume.

For irradiation fluences of~$5\times 10^{14}~\text{n}_\text{eq}/\si{\centi\metre\squared}$ and $1\times 10^{15}~\text{n}_\text{eq}/\si{\centi\metre\squared}$, the CCE obtained via UV-TCT remains lower than for alpha particles, indicating the samples are not fully depleted at the maximum applied reverse bias voltage of 1100~\si{\volt}.
In case of the highest irradiated sample measured, CCE obtained via UV-TCT is slightly higher than for current alpha measurements. The origin of this effect is currently unknown and subject of ongoing studies. The highest obtained CCE values for both measurement types are given in \autoref{tab:comp_UV_alpha}.
Results obtained via UV-TCT approach those using alpha particles with increasing reverse bias voltage.

\begin{table}[ht]
\centering
\caption{Comparison of charge collection efficiencies (CCE) measured using a \textsuperscript{241}Am alpha source and a 370~\si{\nano\metre} UV-TCT setup.}
\label{tab:comp_UV_alpha}
\begin{tabular}{r|l|l}
\textbf{Irradiation Fluence}  & \textbf{CCE [\%]} & \textbf{CCE [\%]} \\ 
\textbf{[$\text{n}_\text{eq}/\text{cm}^2]$} & \textbf{\textsuperscript{241}Am-$\alpha$} & \textbf{UV-TCT \cite{Gaggl_2022}} \\ \hline
$5\times 10^{14}$ & \num{64(11)} & \num{63(3)} \\
$1\times 10^{15}$ & \num{51(10)} & \num{46(3)} \\
$5\times 10^{15}$ & \num{15(3)}\textsuperscript{1} & \num{18(2)}\textsuperscript{1}
\end{tabular}
\end{table}
\vspace{-0.5cm}

\FloatBarrier
\section{Discussion}
The performance of four different planar 4H-SiC diodes neutron-irradiated at fluences between $5\times 10^{14}~\text{n}_\text{eq}/\si{\centi\metre\squared}$ and $1\times 10^{16}~\text{n}_\text{eq}/\si{\centi\metre\squared}$ has been studied using alpha particles from an \textsuperscript{241}Am source.
For the sample irradiated at $1\times 10^{16}~\text{n}_\text{eq}/\si{\centi\metre\squared}$, no signals were able to be acquired.
For the other samples, a decrease in the charge collection efficiency (CCE) was observed, with values up to \SI{64}{\percent}, \SI{51}{\percent} and \SI{15}{\percent} for fluences of $5\times 10^{14}$, $1\times 10^{15}$ and  $5\times 10^{15}~\text{n}_\text{eq}/\si{\centi\metre\squared}$.
For all irradiated samples the CCE increases with the reverse bias voltage, indicating that charge losses due to trapping can be mitigated by increasing charge carrier velocity closer to the saturation velocity.

Comparing these results with UV-TCT measurements~\cite{Gaggl_2022}, a saturation of the CCE at lower bias voltages can be observed, which is explained by the small penetration depth of the alpha particles in the detector. 
At the highest applied bias, the measurements using alpha particles and UV-TCT agree well with each other.
Studies on similar neutron-irradiated 30~\si{\micro\metre} 4H-SiC samples report higher CCEs (up to 55\% and 78\% for neutron fluences of $5\times 10^{14}~\text{n}_\text{eq}/\si{\centi\metre\squared}$ and $1\times 10^{15}~\text{n}_\text{eq}/\si{\centi\metre\squared}$ respectively)~\cite{paper_samples}.
This discrepancy can be explained by different doping concentrations of the samples used (\SI{1.5e14}{\per \centi \meter^3} vs \SI{1.5e15}{\per \centi \meter^3}~\cite{paper_samples}).

The high atomic displacement threshold energy of 4H-SiC suggests enhanced irradiation hardness~\cite{book_fundamentals_of_SiC}. However, the measured radiation hardness proves inferior to comparable epitaxial p-on-n Si samples where the CCE reaches 67.7\% at a neutron irradiation of $2.55\times 10^{15}~\text{n}_\text{eq}/\si{\centi\metre\squared}$, albeit measured at a temperature between \SI{-20}{\celsius} and \SI{-30}{\celsius}~\cite{CCE_Si_with_o_on_n}. One reason might be the high impurity of the employed substrate, about two orders of magnitude higher than for typical Si tracking detectors. Future availability of high resistivity 4H-SiC would enable larger carrier mobility at lower bias. Focusing on electrons for signal creation via the means of n-side readout or the usage of p-substrate diodes could yield a similar effect due to the major increase in charge carrier mobility~\cite{book_fundamentals_of_SiC}. To further improve radiation hardness, understanding the defect creation mechanisms in 4H-SiC naturally is of great importance. In particular, TSC and DLTS measurements can give insight into such processes within the more complex crystal structure. Moreover, detailed characterization of the $\text{SiO}_2/\text{SiC}$ interface is necessary to tackle reliability issues of oxide interfaces, especially under high temperature and radiation environments~\cite{rad_hard_oxide}.

Multiple improvements to the current experimental setup are foreseen:
First, a vacuum setup is currently in construction to remove the air scattering contribution to the energy uncertainty.
Additionally, an alpha source without any layers on top of the isotope should be used in order to obtain a narrower energy spectrum.
In order to investigate the discrepancies between GATE simulations and the measurements, a comparison measurement using a spectroscopy setup (composed of a shaping amplifier and a multi-channel analyzer) instead of high-bandwidth electronics is underway.
Current work in progress also includes the design and development of 4H-SiC-LGADs (low gain avalanche detectors) as a mean to overcome low charge carrier generation rates in 4H-SiC devices.
As the signal generation in LGADs is strongly affected by minor changes in the highly doped gain layer, better understanding of irradiation induced defect creation and donor removal is required.
In order to further characterize the radiation damage in the samples used for this work, doping profile measurements are planned to be conducted.

\paragraph{Funding:}
This project has received funding from the Austrian Research Promotion Agency FFG, grant number 883652. Production and development of the 4H-SiC samples was supported by the Spanish State Research Agency (AEI) and the European Regional Development Fund (ERDF), ref. RTC-2017-6369-3.

\bibliographystyle{JHEP}
\bibliography{iWoRiD}
\end{document}